# Digital Twin: Enabling Technologies, Challenges and Open Research

AIDAN FULLER[1], (Student Member, IEEE), ZHONG FAN[1], CHARLES DAY[1], (Member, IEEE), AND CHRIS BARLOW [2]
[1]School of Computing and Mathematics, Keele University, Staffordshire, ST5 5BG. e-mail: (a.fuller@keele.ac.uk, z.fan@keele.ac.uk, c.r.day@keele.ac.uk).
[2]Astec IT Solutions Limited, Rugeley, Staffordshire, WS15 1UW. e-mail: (cbarlow@astecsolutions.com).
Corresponding author: Aidan Fuller (e-mail: a.fuller@keele.ac.uk)

**ABSTRACT** Digital Twin technology is an emerging concept that has become the centre of attention for industry and, in more recent years, academia. The advancements in industry 4.0 concepts have facilitated its growth, particularly in the manufacturing industry. The Digital Twin is defined extensively but is best described as the effortless integration of data between a physical and virtual machine in either direction. The challenges, applications, and enabling technologies for Artificial Intelligence, Internet of Things (IoT) and Digital Twins are presented. A review of publications relating to Digital Twins is performed, producing a categorical review of recent papers. The review has categorised them by research areas: manufacturing, healthcare and smart cities, discussing a range of papers that reflect these areas and the current state of research. The paper provides an assessment of the enabling technologies, challenges and open research for Digital Twins.

## I. INTRODUCTION

Digital Twin is at the forefront of the Industry 4.0 revolution facilitated through advanced data analytics and the Internet of Things (IoT) connectivity. IoT has increased the volume of data usable from manufacturing, healthcare, and smart city environments. The IoT's rich environment, coupled with data analytics, provides an essential resource for predictive maintenance and fault detection to name but two and also the future health of manufacturing processes and smart city developments [1], while also aiding anomaly detection in patient care, fault detection and traffic management in a smart city [2] [3]. The Digital Twin can tackle the challenge of seamless integration between IoT and data analytics through the creation of a connected physical and virtual twin (Digital Twin). A Digital Twin environment allows for rapid analysis and real-time decisions made through accurate analytics. This paper provides a comprehensive review of Digital Twin use, its enabling technologies, challenges and open research for healthcare, manufacturing and smart city environments. Since the centre of gravity of the literature relates to manufacturing application, the review has tried to capture relevant publication from 2015 onwards across three areas: manufacturing, healthcare and smart cities. The paper, uses a range of academic sources found through keywords related to IoT and data analytics, but with an overall aim of identifying papers relating to Digital Twin.

In conducting this review, we are attempting to answer the following research questions:

RQ1. What is a Digital Twin and what are some of its misconceptions with current and previous definitions?

RQ2. What are the applications, challenges, and enabling technologies associated with IoT/Industrial IoT(IIoT), data analytics and Digital Twins?

RQ3. Is there a link between IoT, IIoT and data analytics with Digital Twin technology?

RQ4. What are the open research and challenges with Digital Twins?

This paper focusses on the status of Digital Twins with IoT/IIoT and data analytics identified as enabling technologies. The rest of the paper is organised as follows: Section II will define a Digital Twin, identifying similar concepts and applications, while highlighting the misconceptions seen in such definitions. Section III discusses the challenges found. Section IV investigates the key enabling technologies for Digital Twins while giving a brief history of each key enabling technologies. Section V relates to current research and is split into three subsections. Subsections A and B set out the methodology for producing the categorical review in Table 5, subsection C follows with a concise analysis of a range of papers on Digital Twins across a plethora of disciplines and finally, the concluding section gives an insight from an industry perspective. Section VI presents open research, with overall challenges and findings for Digital Twin research. Section VII concludes the paper.

## II. DIGITAL TWIN
### A. WHAT IS A DIGITAL TWIN?

The origins of the Digital Twin are set out in this section. The review sets out clear definitions while also looking at some of the misconceptions found with wrongly identified Digital Twins.

Formal ideas around Digital Twins have been around since the early 2000s [4]. That said, it may have been possible to define Digital Twins earlier owing to the ever-changing definitions.

#### 1) Definitions

The first terminology was given by Grieves in a 2003 presentation and later documented in a white paper setting a foundation for the developments of Digital Twins [4]. The National Aeronautical Space Administration (NASA) released a paper in 2012 entitled "The Digital Twin Paradigm for Future NASA and U.S. Air Force Vehicles", setting a key milestone for defining Digital Twins.

a) Nasa 2012 [5]

"A Digital Twin is an integrated multiphysics, multiscale, probabilistic simulation of an as-built vehicle or system that uses the best available physical models, sensor updates, fleet history, etc., to mirror the life of its corresponding flying twin." [5]

b) Chen 2017 [6]

"A digital twin is a computerized model of a physical device or system that represents all functional features and links with the working elements."

c) Liu et al. 2018 [7]

"The digital twin is actually a living model of the physical asset or system, which continually adapts to operational changes based on the collected online data and information, and can forecast the future of the corresponding physical counterpart."

d) Zheng et al. 2018 [8]

"A Digital Twin is a set of virtual information that fully describes a potential or actual physical production from the micro atomic level to the macro geometrical level."

e) Vrabic et al. 2018 [9]

"A digital twin is a digital representation of a physical item or assembly using integrated simulations and service data. The digital representation holds information from multiple sources across the product life cycle. This information is continuously updated and is visualised in a variety of ways to predict current and future conditions, in both design and operational environments, to enhance decision making."

f) Mandi 2019 [10]

"A Digital Twin is a virtual instance of a physical system (twin) that is continually updated with the latter's performance, maintenance, and health status data throughout the physical system's life cycle." [10]

Definition a) is an ambiguous definition specific for NASA's interplanetary vehicle development [5] and is one of the early papers that defines Digital Twins. Despite there being over six years between publications a) and f), the consensus remains that there is not a fundamental or meaningful change. Academia and industry alike have not helped in distinguishing DT's from general computing models and simulations. Future work requires a more definitive definition for a Digital Twin. This research aims to aid in the development of an updated definition, while also helping in analysing related work and pointing out wrongly identified Digital Twins.

### B. DIGITAL TWIN MISCONCEPTIONS
#### 1) Digital Model

A digital model is described as a digital version of a preexisting or planned physical object, to correctly define a digital model there is to be no automatic data exchange between the physical model and digital model. Examples of a digital model could be but not limited to plans for buildings, product designs and development. The important defining feature is there is no form of automatic data exchange between the physical system and digital model. This means once the digital model is created a change made to the physical object has no impact on the digital model either way. Figure 1. illustrates a Digital Model.

#### 2) Digital Shadow

A digital shadow is a digital representation of an object that has a one-way flow between the physical and digital object. A change in the state of the physical object leads to a change in the digital object and not vice versus. Figure 1. illustrates a Digital Shadow.

#### 3) Digital Twin

If the data flows between an existing physical object and a digital object, and they are fully integrated in both directions, this constituted the reference "Digital Twin". A change made to the physical object automatically leads to a change in the digital object and vice versa. Figure 1. illustrates a Digital Twin.

These three definitions help to identify the common misconceptions seen in the literature. However, there are several misconceptions seen but they are not limited to just these specific examples. Amongst the misconceptions is the misconception Digital Twins have to be an exact 3D model of a physical thing. On the other hand, some individuals that think a Digital Twin is just a 3D model.

Figure 1. and their definitions present the different levels of integration for a Digital Twin. Table 5 in Section V of

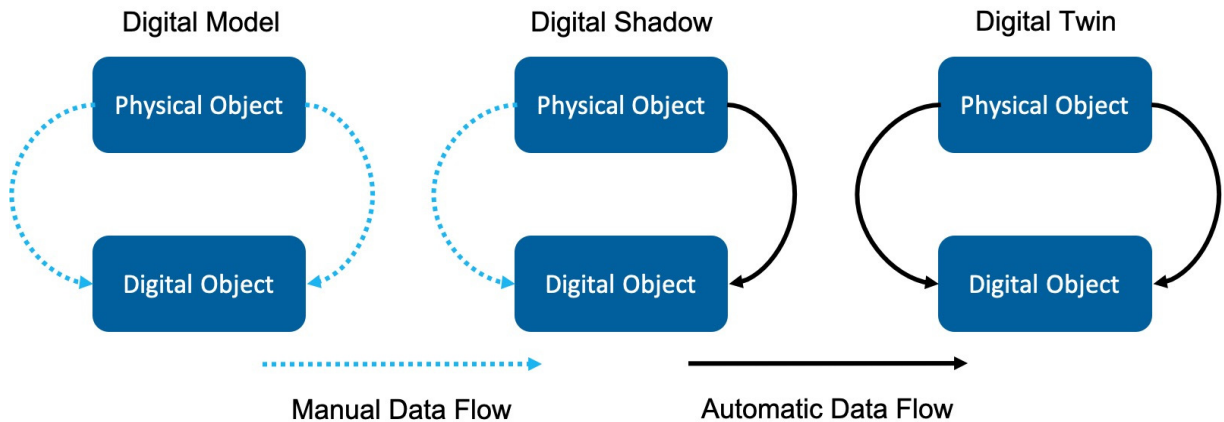

**Figure 1.** Digital Model, Shadow and Twin.

this review presents a range of publications, highlighting the claimed level of integration against the actual integration based on the above definition. The definitions and figures should help in the development and identification of future Digital Twins.

### C. DIGITAL TWIN APPLICATIONS

The next part of this review focusses on the applications of Digital Twins. It will first start by looking at the potential applications for Digital Twins, discussing the domain, sectors, and specific problems for Digital Twin technology. For the moment the term and concept of a Digital Twin are growing across academia, and the advancements in IoT and artificial intelligence (AI) are enabling this growth to increase [11] [12] [13] [14] [15] [16]. At this stage, the primary areas of interest are smart cities and manufacturing with some healthcare-related applications of Digital Twin technology found.

#### 1) Smart cities

The use and the potential for Digital Twins to be dramatically effective within a smart city is increasing year on year due to rapid developments in connectivity through IoT.

With an increasing number of smart cities developed, the more connected communities are, with this comes more Digital Twins use. Not only this, the more data we gather from IoT sensors embedded into our core services within a city, but it will also pave the way for research aimed at the creation of advanced AI algorithms [3] [17] [18].

The ability of services and infrastructures within a smart city to have sensors and to be monitored with IoT devices is of great value for all kinds of future-proofing. It can be used to help in the planning and development of current smart cities and help with the ongoing developments of other smart cities. As well as the benefits of planning, there are also benefits within the energy saving world. This data gives an excellent insight into how our utilities are being distributed and used. Advancement for the smart city is the potential to utilise Digital Twin technology. It can facilitate growth by being able to create a living testbed within a virtual twin that can achieve two things; one, to test scenarios, and, two, to allow for Digital Twins to learn from the environment by analysing changes in the data collected. The data collected can be used for data analytics and monitoring. The scope for Digital Twins is becoming more viable as the development of smart cities increases connectivity and the amount of usable data [19] [20] [21] [22].

#### 2) Manufacturing

The next identified application for Digital Twin is within a manufacturing setting. The biggest reason for this is that manufacturers are always looking for a way in which products can be tracked and monitored in an attempt to save time and money, a key driver and motivation for any manufacturer. Thus why Digital Twins look to be making the most significant impact within this setting. Likewise, with the development of a smart city, connectivity is one of the biggest drivers for manufacturing to utilise Digital Twins. The current growth is in line with the Industry 4.0 concept, coined the 4th industrial revolution, this harnesses the connectivity of devices to make the concept of Digital Twin a reality for manufacturing processes [23] [1] [24] [25] [26].

The Digital Twin has the potential to give real-time status on machines performance as well as production line feedback. It gives the manufacturer the ability to predict issues sooner. Digital Twin use increases connectivity and feedback between devices, in turn, improving reliability and performance. AI algorithms coupled Digital Twins have the potential for greater accuracy as the machine can hold large amounts of data, needed for performance and prediction analysis. The Digital Twin is creating an environment to test products as well as a system that acts on real-time data, within a manufacturing setting this has the potential to be a hugely valuable asset [27] [28] [2] [29].

Another application of Digital Twins is in the automotive industry, most notably demonstrated by Tesla. The ability to have a Digital Twin of an engine or car part can be valuable in terms of using the twin for simulation and data analytics [30] [31]. AI improves the accuracy of testing as it can perform data analytics on live vehicle data to predict the current and future performance of components.

The construction industry is another sector that hosts a range of applications for Digital Twin use. The development stage of a building or structure is a potential application for a Digital Twin. The technology cannot only be applied in the development of smart city buildings or structures but also as an ongoing real-time prediction and monitoring tool. The use of the Digital Twin and data analytics will potentially provide greater accuracy when predicting and maintaining buildings and structures with any changes made virtually then applied physically. The Digital Twin gives construction teams greater accuracy when carrying out simulations as the algorithms can be applied in real-time within the Digital Twin before the physical building.

A common goal seen so far across the field of Digital Twins is this idea of real-time simulation as opposed to low detailed static blueprint models. The use of these models serves a purpose, but they are not using real-time parameters which limit the predictability and learnability. The Digital Twin can be learning and monitoring simultaneously, as well as applying machine and deep learning algorithms [32] [33] [34] [35].

3) Healthcare

The healthcare sector is another area for the application of Digital Twin technology. The growth and developments enabling technology are having on healthcare is unprecedented as the once impossible is becoming possible. In terms of IoT the devices are cheaper and easier to implement, hence the rise in connectivity [36] [37]. The increased connectivity is only growing the potential application of Digital Twin use within the healthcare sector. One future application is a Digital Twin of a human, giving a real-time analysis of the body. A more realistic current application is a Digital Twin used for simulating the effects of certain drugs. Another application sees the use of a Digital Twin for planning and performing surgical procedures [38].

Likewise with other applications within a healthcare setting the use of a Digital Twin gives researchers, doctors, hospitals and healthcare providers the ability to simulate environments specific to their needs whether it be real-time or looking to future developments and uses. As well as this, the Digital Twin can be used simultaneously with AI algorithms to make smarter predictions and decisions. Many applications within healthcare do not directly include the patient but are beneficial for the ongoing care and treatment, hence the key role such systems have on patient care. Digital Twin for healthcare is in its infancy, but the potential is vast from using it for bed management to large scale wards and hospital management.

Having the ability to simulate and act in real-time is even more paramount within healthcare as it can be the difference between life or death. The Digital Twin could also assist with predictive maintenance and ongoing repair of medical equipment. The Digital Twin within the medical environment has the potential along with AI to make life saving decisions based on real-time and historical data [39] [40].

Applications of a Digital Twin are identified here, showing some of the cross overs in the intended use demonstrating how predictive maintenance is adaptable from manufacturing plant machines to patient care. It also shows some of the applications where they do not cross over, and Digital Twin use is specific to its intended use. The advancements in AI, IoT and Industry 4.0 have facilitated the growth in Digital Twin applications.

### D. DIGITAL TWIN IN INDUSTRY

General Electric (GE) first documented its use of a Digital Twin in a patent application in 2016. From the concept set out in the patent, they developed an application called the "Predix" platform [41] which is a tool for creating Digital Twins. Predix [41] is used to run data analytics and monitoring. In recent years, GE has scaled back their plans for a Digital Twin, planning to focus on their heritage as an industrial multinational rather than a software company. Siemens, however, has developed a platform called "MindSphere" [42] which has embraced the Industrial 4.0 concept with a cloud based system that connects machines and physical infrastructure to a Digital Twin. It uses all the connected devices and billions of data streams with the hope of transforming businesses and providing Digital Twin solutions [42].

An alternative platform for developing Digital Twin and AI technology is "ThingWorx" [22]. This platform created by PTC is an Industrial Innovation Platform with the main focus of harvesting IIoT/IoT data and presenting via an intuitive, role-based user interface that delivers valuable insight to users. The platform facilitates the smooth development of data analytics while also developing an environment for a Digital Twin solution [22].

IBM developed a platform called "Watson IoT Platform" [43] marketed as an all-round IoT data tool that can be used to manage large scale systems, in real-time, through data collected from millions of IoT devices. The platform has several add on features: cloud based services, data analytics, edge capabilities and blockchain services. All of which makes this a possible platform for a Digital Twin system [43].

From an open-source viewpoint, there are two big projects to highlight. The first is the "Ditto" project by Eclipse [44], a ready to use platform that can manage the states of a Digital Twin, giving access and control to physical and Digital Twins. The platform lies in a back-end role providing support for already connected devices and simplifying the connection and management of Digital Twins [44]. Another open-source project called "imodel.js" developed by Bentley Systems [45] is a platform for creating, accessing and building Digital Twins.

## III. CHALLENGES

It is becoming more evident that Digital Twin runs in parallel with AI and IoT technology resulting in shared challenges. The first step in tackling the challenges is to identify them. Some of the common challenges are found with both data analytics and the Internet of Things, and the end aim is to identify shared challenges for Digital Twins.

### A. DATA ANALYTIC CHALLENGES

Some of the challenges within the field of machine and deep learning are listed below.

#### 1) IT Infrastructure

The first big challenge is the general IT infrastructure. The rapid growth of AI needs to be met with high-performance infrastructure in the form of up to date hardware and software, to help execute the algorithms. The challenge with the infrastructure currently is down to the cost of installing and running these systems. For instance, the costs of the high-performance graphics processing unit (GPUs) that can perform the machine and deep learning algorithms are in the thousands, anything from \$1,000 to \$10,000. As well as this, the infrastructure needs updated software and hardware to run such systems successfully. Overcoming this challenge is seen through the use of GPUs "as a service" providing on-demand GPUs at cost through the cloud. Amazon, Google, Microsoft and NVIDIA, to name a few, are offering unique on-demand services similar to traditional cloud-based applications, breaking the barrier to demand, but the poor infrastructure and high cost are still challenging for data analytics. Using the cloud for data analytics and Digital Twins still pose challenges in ensuring that the cloud infrastructure offers robust security.

#### 2) Data

From a data point of view, it is important to ensure it is not of inferior quality. The data needs to be sorted and cleaned, thereby ensuring the highest quality of data is fed into the AI algorithms.

#### 3) Privacy and Security

Privacy and security is an important topic for anyone concerned with the computing industry and this is no different when performing data analytics. Laws and regulation are yet to be established fully because of the infancy of AI. The challenge is more scrutiny, regulation and measures concerning AI in the future as the technology grows. Future regulation ensures the development of algorithms that take steps to protect user data. The General Data Protection Regulation (GDPR) is a new regulation that ensures the privacy and security of personal data across the UK and throughout Europe. Despite being an umbrella regulation concerning data and security, this highlights the concerns with handling data when developing AI algorithms.

Regulation is one step to ensure personal data is protected, while another method is federated learning, a decentralised framework for training models. It allows users' data in a learning model to stay localised without any data sharing, addressing privacy and security issues when implementing data analytics within a Digital Twin.

#### 4) Trust

Trust is another challenge that concerns much of the field of AI. Firstly, being because it is relatively new and secondly because unless the developer is familiar with the complexity, the use of AI can be daunting. The anxiety that robots and AI will become a dominant force on earth, taking control of key infrastructure from humans is a barrier to trust.

The issue of trust can be a barrier because the portrayal of the AI mostly focuses on the negative effects that could occur. Positive media stories in the field of artificial intelligence are becoming more common, but the challenge is evident, and the need for wider exposure of AI and the positive uses would help overcome challenges with trust. Privacy and security challenges contribute to these trust issues, but more comprehensive privacy and security regulation in AI builds trust.

#### 5) Expectations

The last challenge for data analytics is the expectation that it can be used to solve all our problems. Careful consideration is vital for AI use and investing time in this identifies the correct application, ensuring standard models could not produce the same results. The same as other new technologies, they have the potential to work hand-in-hand with strengthening things like manufacturing and smart city developments.

The potential users only see the benefits and believe it will instantly save time and money, hence the high expectations. The field is still in its infancy, and the challenge needs to be kept in mind when applying data analytics. It is evident through the number of scenarios that use "AI" for processes that do not need it, in contrast to other situations where AI should be used. Greater exposure and understanding of AI is needed to allow people to gain the correct baseline knowledge of the area, thus learning how it can be applied.

### B. IOT/IIOT CHALLENGES

Listed below are the challenges found in the field of internet of things and industrial internet of things:

#### 1) Data, Privacy, Security and Trust

With the huge growth of IoT devices both in the home and industrial setting comes the challenge of collecting substantial amounts of data. The challenge is trying to control the flow of data, ensuring it can be organised and used effectively. The challenge becomes a bigger problem with the advent of big data. The use of IoT increases the large volumes of unstructured data. For IoT to manage the amount of data, sorting and organisation of data is a necessity and will result in more data being usable and providing value. Otherwise, the data collected through IoT will be lost or it will be too cost-prohibitive to extract the value from the enormous volumes amassed.

As the data could be sensitive, it could be of value to a criminal, thereby increasing the threat. The threat is significantly increased for businesses when they could be dealing with sensitive customer data. Cyber-attacks pose more challenges with criminals targeting systems and taking them offline, to cripple an organisation's infrastructure. Some organisations have thousands of connected IoT devices posing a risk that cyber-criminals may target them to take control and use the devices for their services. An example of this is the Mirai botnet scandal were nearly 15 million IoT devices worldwide were compromised and used to launch a distributed denial-of-service attack (DDoS) [46]. The risk of DDoS attacks increases because of the rapid growth of IoT. As well as this, the lag in priorities around privacy and security solutions poses a further risk of attack. When installing the devices, the most up to date security features and protection are needed, if not this is a vulnerability which offers a back door for criminals to infiltrate a larger connected IoT environment.

2) Infrastructure

The IT infrastructure currently in place is behind, due to the rapid growth observed in IoT technology compared with the existing systems currently in place. The updating of old infrastructure and the integration of new technology helps facilitate IoT growth.

Updated IoT infrastructure provides an opportunity to benefit from the latest technology and leverage the applications and services available in the cloud without expensive refreshing of existing systems and technology.

Another challenge for IoT systems is connecting old machines to the IoT environment. One of the ways to combat this is retrofitting IoT sensors to legacy machines, ensuring data is not wasted and old machines can have some form of analytics.

3) Connectivity

Despite this growth in IoT use, the challenges of connectivity still exist. These are especially prevalent when trying to achieve the goal of real-time monitoring. A large number of sensors within one manufacturing process poses a significant challenge when trying to connect all of them simultaneously.

Challenges with attributes like power outages, software errors or ongoing deployment errors are impacting this overall goal of connectivity. Just having one sensor not fully connected could dramatically affect the overall goal of a given process. For example, IoT devices are one source of feeding data to AI algorithms; this can become a major challenge as all the data is required for it to perform accurately and missing IoT data could detrimentally affect the running of the system. Retrofitting machines and harvesting the data already served up by the machine is a method of ensuring all data is collected. Imputation methods are a process of finding replacement values for missing IoT sensor data, a concept used to ensure full connectivity and facilitate the running of AI models with high accuracy and little to no missing data.

4) Expectations

Likewise, with AI, the expectations associated with IoT are a challenge, due to organisation and end-users not fully understanding what to expect from IoT solutions or how to best use them. A promising aspect is that the rapid growth in IoT indicates the end-users and organisations recognise the value in IoT and how a smarter connected world can benefit us all.

The expectation that IoT can just be used infinitely without prior knowledge can be damaging, with the knock-on effect, posing more pressure on privacy and security concerns further putting the burden on challenges with trust. Similar to AI, background knowledge in IoT is needed to ensure it is used to its full potential.

### C. DIGITAL TWIN CHALLENGES

This section draws primarily on the challenges associated with Digital Twins. However, as the research progresses, it is clear to see the challenges found in data analytics, IoT and IIoT are similar to those found in the challenges for Digital Twins with some discussed below:

1) IT Infrastructure

Similarly to both analytics and IoT the challenge is with the current IT infrastructure. The Digital Twin needs infrastructure that allows for the success of IoT and data analytics; these will facilitate the effective running of a Digital Twin. Without a connected and well thought through IT infrastructure, the Digital Twin will fail to be effective at achieving its set out goals.

2) Useful Data

The next challenge is around the data needed for a Digital Twin. It needs to be quality data that is noise-free with a constant, uninterrupted data stream. If the data is poor and inconsistent, it runs the risk of the Digital Twin underperforming as it's acting on poor and missing data. The quality and number of IoT signals is an essential factor for Digital Twin data. Planning and analysis of device use are needed to identify the right data is collected and used for efficient use of a Digital Twin.

3) Privacy and Security

Within an industry setting, it is clear that the privacy and security associated with Digital Twins are a challenge. Firstly because of the vast amount of data they use and secondly the risk this poses to sensitive system data. To overcome this challenge, the key enabling technologies for Digital Twins - data analytics and IoT - must follow the current practices and updates in security and privacy regulations. Security and privacy consideration for Digital Twins data contribute to tackling trust issues with Digital Twins.

### 4) Trust

The challenges associated with trust are both from an organisation point of view and that of the user. Digital Twin technology needs to be discussed further and explained at a foundation level to ensure the end-users and organisations know the benefit of a Digital Twin, which will aim to overcome the challenge of trust.

Model validation is another way to overcome the challenges with trust. Verifying that Digital Twins are performing as expected is key for ensuring user trust.

With more understanding, trust in Digital Twins prevails. The enabling technology will give more insight into the steps that ensure privacy and security practices are followed through development, in turn, overcoming challenges with trust.

### 5) Expectations

Despite Digital Twin adoption being accelerated by industry leaders Siemens and GE, caution is needed to highlight the challenges that exist for the expectations of Digital Twins and the need for more understanding. The need for solid foundations for IoT infrastructure and a greater understanding of data required to perform analytics will ensure the organisations will make use of Digital Twin technology. It is also a challenge to combat the thinking that the Digital Twin should be used solely because of the current trends. The positives and negatives for the expectation of Digital Twins need to be discussed to ensure appropriate action when developing Digital Twin systems.

It is clear to see that challenges for both the Industrial IoT/IoT and data analytics are also shared challenges for the application of a Digital Twin. Despite the challenges Digital Twin shares with IoT and data analytics from a user perspective to the privacy and infrastructure challenges of Digital Twin, there are also specific challenges relating to the modelling and building of the Digital Twin.

### 6) Standardised Modelling

The next challenges within all forms of a Digital Twin development relates to the modelling of such systems because there is no standardised approach to modelling. From initial design to a simulation of a Digital Twin there needs to be a standard approach, whether it be physics-based or designed based. Standardised approaches ensure domain and user understanding while ensuring information flow between each stage of the development and implementation of a Digital Twin.

### 7) Domain Modelling

Another challenge as a result of the need for standardised use is related to ensuring information relating to the domain use is transferred to each of the development and functional stages of the modelling of a Digital Twin. This ensures compatibility with domains such as IoT and data analytics, allowing for the successful uses of the Digital Twin in the future.

These are important moving forward as it ensures they are considered in the future development of Digital Twins as well as when using IIoT/IoT and data analytics. Table 1 below shows a summary of challenges for both data analytics and I/IoT while showing the overarching combined challenges for a Digital Twin, with challenge six and seven specifically for Digital Twin implementation.

**Table 1.** Shared Challenges.

| Digital Twin | |
|---|---|
| *Data Analytics* | *Industrial IoT/IoT* |
| IT Infrastructure | IT Infrastructure |
| Data | Data |
| Privacy | Privacy |
| Security | Security |
| Trust | Trust |
| Expectations | Expectations |
| | Connectivity |

## IV. ENABLING TECHNOLOGIES

This section discusses the enabling technologies for Digital Twins.

### A. BRIEF HISTORY OF THE INTERNET OF THINGS

The Internet of Things is the term given to devices connected to the internet. It is about giving so-called "things" a sense of intelligence and the ability to collect information on their environment. The term was first published in the late 1990s with Kevin Ashton setting out his vision for IoT [47]. The idea that all devices that are interconnected gives the developer the ability to track and monitor everything we do, thus leading to a smarter world. An example of this is to be found many years earlier at Carnegie Mellon University in Pittsburgh. Here a programme would connect a Coca-Cola machine via the Internet to see if the drink was ready and cooled enough for a user to buy and consume [48]: a simple but effective use case for Ashton's vision.

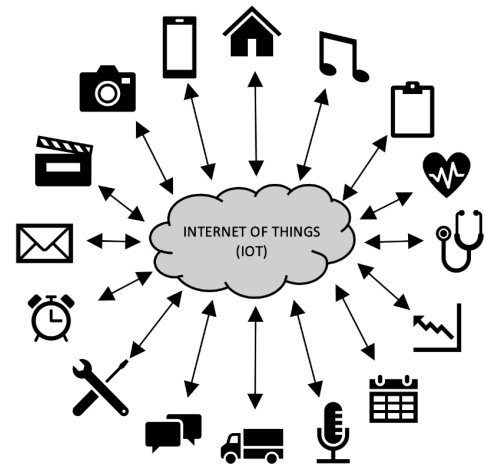

**Figure 2.** Internet of Things Diagram.

The number of IoT devices recorded year on year shows the considerable growth of this technology. In 2018 the figure was over 17 billion [49]. By the year 2025, [50] predicts that there will be over 75 billion devices with the industry predicted to be worth over $5 trillion [51]. Figure 3. shows the growth in IoT devices since 2016. These figures show the enormous impact these devices are having and further adds to the vision set out by Ashton. The considerable number of connected devices aids the vision of a fully connected world, Figure 2. illustrates this idea of connected a services through IoT. The proliferation of IoT devices is universally beneficial, impacting the core of daily life, the communication sector, healthcare, building and transport, smart cities and manufacturing [51] [52].

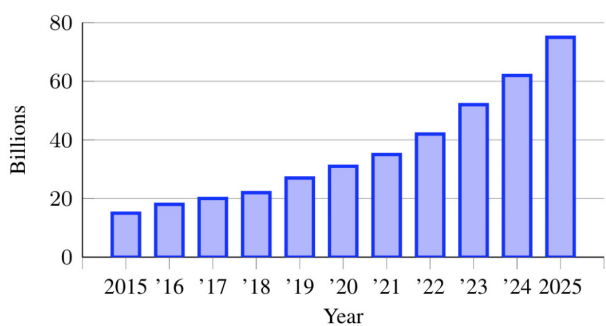

Figure 3. IoT Device Growth. [50]

### B. BRIEF HISTORY OF THE INDUSTRIAL INTERNET OF THINGS

The concept of the Industrial Internet of Things (IIoT) has come from the term IoT, drafted by Ashton [47]. The definition of IoT varies across academia, and the same goes for defining IIoT. The term is similar in characteristics to IoT but with an added emphasis on industrial processes. Boyes et al. present a range of definitions for IIoT, but the main focus outlined is improving productivity for industry [53]. Within manufacturing and industrial settings, the original systems are Industrial Control Systems (ICS). These are well documented and used, but the benefits of these systems becoming autonomous and smart are potentially seen through IIoT. Another technology intricately linked to both IoT and IIoT is Cyber- Physical Systems (CPS) [54] [55]. Both ICS and CPS are like IIoT, but not the same. The main difference being IIoT devices require a connection to the internet as opposed to being enclosed in an ICS architecture [56].

Like IoT, IIoT can have a huge impact on improving manufacturing processes, allowing for tasks to be evaluated with greater knowledge and real-time responses through connected devices, thus improving the performance, production rate, costs, waste and many other critical deliverables within the industry setting [57]. The IIoT does not only affect manufacturing but agriculture, oil, gas and other large scale processes. Likewise, with IoT, Industrial IoT is having a significant impact within the industry. This is especially seen with Morgan Stanley predicting the market size to reach $110 billion by 2021 [58]. [59] reports that IIoT could add $14.5 trillion to the global economy by 2030 [59]. Figure 4. shows the development of Industry 4.0, which is the introduction of IIoT within the industrial revolution timeline [60].

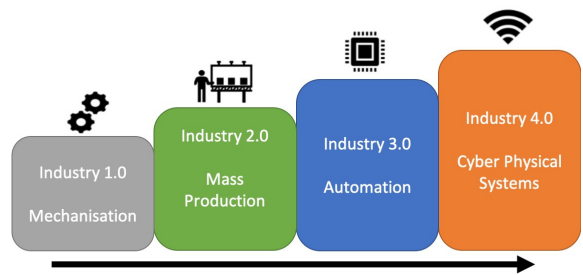

Figure 4. Industrial Revolution.

### C. I/IOT ENABLING TECHNOLOGIES AND FUNCTIONAL BLOCKS

Both IoT and IIoT have a wide range of essential areas that ensure the running of connected systems. These enabling technologies are classified into four main functional domains, as described by [61]. These domains cover the individual enabling technologies from network communication, hardware and software to data processing, power and energy storage — all with specific goals to enable the full development of an IoT system facilitating an Industry 4.0 architecture.

The four enabling technology domains for I/IoT comprise of D1 the Application domain, D2 the Middleware domain, D3 the Network domain and D4 the Object domain as seen below in Table 2.

Table 2. Enabling Technologies and Functional Blocks: I/IoT.

| Domain | Enabling Technology |
|---|---|
| D1 Application Domain | I/IoT Application |
| | Architecture |
| | Software and APIs |
| D2 Middleware Domain | Cloud Platforms |
| | Data Processing Mechanism |
| | Data Storage |
| D3 Networking Domain | Communication Protocol |
| | Network Interface |
| | Adoption Mechanism |
| D4 Object Domain | Hardware Platform |
| | Embedded Objects |
| | Mechanical and Electrical Parts |

D1 is made up of three layers. The first is the application layer, which is the I/IoT applications; from smart home and smart cities to smart farms. Next is the architecture layer; this can be enabling software architectures; SOA (Service-oriented architecture) or REST (Representational State Transfer), both examples of what makes up the architecture layer. The third layer; software and APIs, bridges the application domain to the middleware domain. It maintains the operating systems and software. For instance, Android and custom made OS's used to operate an IoT system. This could also be made up of custom-built APIs for the deployment of an IoT system, both of which are a key technology for bridging D1 to D2.

The middleware domain is made up of three more layers. The first being the cloud platform, which is made up of services that provide on-demand computing resources through the cloud. Microsoft, Amazon and Google are leading providers in cloud services.

The second layer is data processing; enabled using data mining and example services provided by BigQuery, Apache and Storm. The third enabling layer in D2 is data storage. This is essential in an I/IoT infrastructure, and an example is MongoDB, which offers large storage engines.

The third part of the IoT system is D3, the networking domain, which is made up of three enabling layers. The first is the communication protocol layer; this comprises of the application, transport and network protocols for a given system, enabling seamless communication. The second enabling layer is the network interface. Located here are essential technology standards (for example RFID) used throughout the IoT system, again for enabling the seamless integration of IoT. The final layer of D3, the networking domain, is the adoption mechanisms. Consisting of the adoption layer, which includes standards like 6TiSCH and IEEE 1095, which enable more reliable wireless communication; likewise with the connectivity interface and the gateway layer, all of which are key enabling technology standards for the development of an I/IoT system.

D4 is the final block in the IoT system, as illustrated in Table 2. The object domain is made up of three enabling layers. The first is the hardware platforms, consisting of the hardware solutions, examples being Raspberry Pi or Arduino. This domain brings together the last three layers; examples being sensors, radio tags, displays and firmware, all of which are vital in connecting the system. The last layer is the mechanical and electrical parts, made up of the batteries and the processing units needed to run the device.

The splitting up of the domain into four functional blocks is easier for understanding the twelve enabling sections of this given IoT system; this provides an integrated and interconnected framework.

### D. BRIEF HISTORY OF DATA ANALYTICS

The term data analytics is an umbrella term that groups analytic concepts, as seen throughout the paper and academia. Therefore an understanding and analysis of other papers are needed. The term data analytics stems from the field of "Data Science", a multidisciplinary subject that covers a range of concepts, with an emphasis on collecting and presenting data for analysis to gain greater insight. The subsection below presents an in-depth analysis of the field of data analytics. The identification and highlighting of these topics will help in analysing other papers and seeing where this research fits in [62].

#### 1) Data

To perform data analysis, the need for raw data is paramount. There are several actions needed to turn this data into usable information, ready for use in algorithms and statistical analysis. These being the requirements, collection, processing and cleaning. The requirements set out the necessary needs of the data and how it is used, ensuring that specific requirements are outlined, considering the intended use of the data. The second stage acts on the requirement of collecting the relevant data, identifying physically where and how the data will be collected. The collected data will then go through a processing phase in which it is sorted according to specific requirements. The final phase and arguably the most important is the cleaning of data. Despite the data being collected and sorted, it may have significant gaps or erroneous data. This cleaning phase uses the imputation methods, previously identified as challenges to data analytics. These methods ensure that no missing data exists [63] [24].

#### 2) Statistics

Statistics is the overarching term for the collection, classification, analysis, and interpretation of data. Briefly relevant in this case for data analysis as statistical models underpin machine learning algorithms. Statically inference and descriptive statistics are another way in which data analytics are used to describe observations in collected data. AI and the following topics below show the growth of advanced data analytics [64] [63].

#### 3) Artificial Intelligence

Artificial Intelligence (AI) is the first topic of interest in data analytics. The overall definition of AI dates back to the late 50s with this concept of creating "intelligent systems" [65]. These are categorised below into topics of potential importance for this project [66] [67].

#### 4) Machine Learning

A subsection of AI, machine learning is the creation of algorithms that can give the computer the ability to learn and act for the user without being directly programmed to do so. Machine learning is used to create programmes that use sophisticated algorithms to collect and analyse data autonomously. For more general analysis, machine learning can fit into two types of learning: [54] [55].

##### a) Supervised Learning

This is the most popular form of machine learning. The algorithms use large amounts of labelled data to analyse and learn. The algorithm is tasked with learning and analysing the labelled data to identify a given task correctly; image classification is one example [68]. The algorithms learn from training data and are then given test data to see how well it is accurately predicting what an image is showing, presented through an accuracy percentage. The user then analyses these answers and any errors are corrected and re-learned, helping train the model and increasing the accuracy of a given algorithm [69].

### b) Unsupervised Learning

Unsupervised learning is another form of machine learning, it does not require expensively marked-up data where for each input pattern the desired output has previously been determined: as is required for supervised learning [68]. Unsupervised learning algorithms learn using it's own methods in categorising and highlighting patterns within data instead of relying on user feedback. Clustering is one method of categorising data. Algorithms learn to cluster unlabelled data sets together, potentially showing hidden patterns that were not explicitly identifiable [68].

### c) Deep Learning

Deep learning is another part of the field of data analytics and a subsection of machine learning. Deep learning algorithms learn unstructured and unlabelled data using complex neural networks with autonomous input feature extraction as opposed to manual extraction [70]. These networks utilise machine learning to create deep learning models that can take longer to train because of the much larger neural networks, but this allows for greater accuracy. Another type of learning is semi-supervised learning, defined as having some labelled data, but more data is unlabelled to see how the algorithms can learn to be more accurate [70]. Many more algorithms appear throughout the field of data science, but these are the most common.

### 5) Data Visualisation

The final subtopic within data analytics is visualisation, defined as a graphical representation or visualisation of data or results. The type of data affects the way it is visualised. The most common being multidimensional data, which can be presented using graphs and charts, taking multiple variables, for instance, bar or pie charts. Another data type is geospatial; this involves data collected from the earth through location data, visualised through distribution maps, cluster maps, and more commonly, contour maps [71].

### E. ENABLING TECHNOLOGIES AND FUNCTIONAL BLOCKS FOR DATA ANALYTICS

The next section concerns data analytics within the field of Artificial Intelligence, machine and deep learning. The descriptions of enabling technologies for data analytics [72] and the classification scheme outlined by [61] are summarised in Table 3 below. The enabling technologies are like IoT in many ways but have slightly different layers around visualisation and the algorithm side of analytics. An overview is seen below, with the domains labelled D5, D6, D7 and D8.

Table 3 is produced as a result of analysing Table 2 by Bibri and Krogstie [72]; however, it is slightly different in presentation. The table starts with D5, the object domain followed by D6, the middleware domain, D7, the network domain, and lastly D8, the application domain. In each of the domains is a list of enabling technologies associated with data analytics.

Table 3. Enabling Technologies and Functional Blocks: Data Analytics.

| Domain | Enabling Technology |
| --- | --- |
| D5 Object Domain | Data Collection and Preprossessing |
|  | Data Repositories |
| D6 Middleware Domain | Storage Facilities |
|  | Data Processing |
|  | Analysis Techniques and Algorithms |
| D7 Networking Domain | Wireless Network Technology |
| D8 Application Domain | Hardware & Data Visualisation |
|  | Data Analytic Applications |

D5 is the object domain which has at least three layers, reflecting the dual-status of the storage facilities. The first enabling layer is the data collection, which deals with the pre-processing of data for the analytic solutions. The use of data sensing tools and methods enables the collection of data. Digital signal processing units also ensure the harvesting of data. The second layer is the data repository which facilitates the storage and use of databases. The final layer of D5, linking to D6, is the storage facilities which enable the storage of copious amounts of data through the use of server storage enabling on-demand data. This layer is also the connection to the processing of the storage data to the middleware domain.

D6 is the middleware domain, consisting of three enabling layers. The first links with D5, which relates to storage processing. The second layer in D6 is data processing, which is the main layer for enabling data analytics, cloud services and the main middleware architectures, including software and database systems. The third layer in D6, seen in Table 2, is the analysis and algorithms. This layer facilitates the task of data mining, machine learning, statistics and querying of the collected data. As well as the enabling models within data analytics; supervised and unsupervised learning.

D7 is the networking domain, showing enabling technology for the connectivity protocols looking at wireless and communication and how they enable efficient collection and processing of data from previous layers and domains. D7 also concerns the enabling standards relating to the privacy and security mechanisms.

The final domain discussed is D8, entitled application. D8 has two enabling technology layers. The first being the hardware and visualisation layers. This layer enables tangible technology to record the data and conduct machine and deep learning or statistical analysis. The visualisation side of the layer easily enables the display of useful information regarding user tasks.

Finally, the application layer highlights the applications relating to data analytics such as self-driving cars, image recognition or virtual personal assistants such as Amazon's Alexa.

In summary table 2 presents the functional domains for enabling technologies associated with IoT/IIoT, while Table 3 presents the functional domains for enabling technologies for data analytics. The above section provides a framework for creating a synthesis of functional domains and enabling technologies in the context of a Digital Twin, as can be seen in Table 4 of subsection F, below.

### F. ENABLING TECHNOLOGIES FOR DIGITAL TWIN

Similar to Table 2 by Bibri and Krogtie [72], Table 4 provides an additional synthesis of ideas for the functional domains and enabling technologies of a Digital Twin. Starting with D9 for the Application domain, progressing to D10 for the Middleware domain, followed by D11 for the Networking domain and finally D12 for the Object domain.

Table 4. Enabling Technologies and Functional Blocks: Digital Twin.

| Domain | Enabling Technology |
|---|---|
| D9 Application Domain | Model Architecture and Visualisation |
|  | Software and APIs |
|  | Data collection and Pre-processing |
| D10 Middleware Domain | Storage Technology |
|  | Data Processing |
| D11 Networking Domain | Communication Technology |
|  | Wireless Communication |
| D12 Object Domain | Hardware Platform |
|  | Sensor Technology |

The first domain is D9, which is the application domain and is made up of three important layers, the first being the model architecture and visualisation layer essential for creating high-fidelity models of the physical entity. The layer enables the visualisation and architecture modelling of Digital Twins. This ensures Digital Twins are modelled using more than just the behaviours of physical entities. It is enabled through tools such as Simulink and Twin Builder. The second layer is the Software and API, specifically used to aid in the modelling of such Digital Twin architecture facilitating the third layer; pre-processing and collection. This last layer of the application domain is needed to ensure the data is collected properly, for example, with Predix, Mindsphere and Storm, to name just a few applications for data collection. This layer ensures the data is harvested correctly to facilitate the use of IoT and analytics for a Digital Twin while also bridging domain D9 to D10.

D10, the middleware domain, consists of two enabling layers. The first being storage technology. Which facilitates the storage of data through Mongo DB, MySQL services and on-demand databases which are needed for Digital Twin use. The second layer, related to data processing, is essential to transfer the stored data between D10 and D11.

D11 consists of the Network Domain with two enabling layers, the first being the communication Technology layer essential in ensuring the data collected is communicated between domains. The second layer in the network domain functional block for Digital Twins is the wireless communication layer, which is needed to ensure the transmission of data wirelessly follows the correct protocol within a Digital Twin architecture as well as bringing data to the next domain, D12.

D12, the object domain, consist of two enabling layers. The first is the hardware platform and the second being the sensor technology. Both are needed to ensure the correct hardware is in place to conduct Digital Twin analysis, as well as facilitating the collection of data through sensor technology.

## V. CURRENT RESEARCH

This next section identifies related work for IoT/IIoT and data analytics with a focus on Digital Twins publications, discussing a range of publications and identifying gaps in the field. This insight from academia will aid other researchers, enabling them to find gaps within the research and will allow a move towards a more comprehensive definition of a Digital Twin.

### A. CATEGORICAL REVIEW METHODOLOGY

The first part of this section performs a categorical review of the relevant literature, following methodology used by Kritzinger et al, [73] to produce a categorical review table of selected publications. The main elements of the review draw on the three levels of integration of a Digital Twin, as described in Section II, subsection B of this paper and by Kritzinger et al. [73].

The methodology used in this paper is based on work conducted by Kritzinger et al. [73] ], which involved the categorisation of forty-three papers related to the topic of Digital Twins, that were published between from 2001 to 2017. The papers were obtained via literature search engines, such as Google Scholar. For the search performed in this paper, the authors used Google Scholar with specific searches targeting ACM, IEEE and Science Direct repositories. Of the research found there were 177 papers to look at from 2015 to present (31st December 2019), with only 42, pre-2017. The search terms included variations of Digital Twin (Digital-Twin, Digital Twins). As well as the term Digital Twin, the search included adding terms relating to the broad research areas (Industrial Digital Twin, Healthcare Digital Twin, Smart Cities Digital Twin). We found twenty-six key sources from three areas, manufacturing, healthcare and smart cities to synthesise Table 5.

### B. ORGANISATION OF THE CATEGORICAL REVIEW

The papers are ordered alphabetically in three broad areas, manufacturing, healthcare and smart cities, shown below in Table 5. The following sections discuss the columns used in Table 5:

#### 1) Paper

The first column provides the authors and year of each publication.

#### 2) Type

The second column identifies each paper by the type of research carried out; review paper or a case study. It could also be a new concept relating to Digital Twins or a definition, each is categorised accordingly.

### 3) Defined Twin

The next two columns are the definition relating to levels of integration. The first identifying what the original authors are referring to; Digital Model, Shadow or Twin.

### 4) Actual Twin

The second column identifies if the paper accurately describes a Digital Model, Shadow or Twin. The definitions from Section II are used to classify the publication, giving insight into possible misconceptions in definitions.

### 5) Broad Area

A fifth column identifies the broad areas of research: this being manufacturing, healthcare and smart cities.

### 6) Specific Area

The sixth column elaborates on the broad area narrowing down to a specific area. For instance, in manufacturing this would be narrowed down to a smart factory or fault diagnosis area of interest. For smart cities, this is narrowed down to traffic or infrastructure.

### 7) Technology

The final column identifies the technology used, for example, simulation, data analytics and IoT.

The papers are ordered alphabetically in three broad areas, manufacturing, healthcare and smart cities, shown below in Table 5.

### C. ANALYSIS OF THE REVIEWED PAPERS

The following sections draw on the papers reviewed in the categorical review, discussing the paper in more detail, highlighting any concepts and case studies performed. The sections are not limited to but will include, the main areas of papers collected relating to healthcare, smart cities and manufacturing. The order of the areas presented reflects the level of current research in terms of the number of papers found. For healthcare, the number of papers found is limited, but the potentially life-changing benefits Digital Twins can have on the healthcare industry are prevalent [84] [32] [39] [40]. Next closely is the smart city area with a small number of papers found. Most research falls within the manufacturing setting. Figure 5. shows the share of papers in terms of the area of research they fit in to; healthcare, manufacturing or smart cities, all discussed below with example papers.

### 1) Healthcare

An element to take from the several definitions of Digital Twins is the concept as described by He [75], the "digital replications" of a physical thing. El Saddik redefined this with the inclusion of Digital Twin replications of living things as well as non-living entities [39]. Presenting potential use of the Digital Twin for the healthcare sector shows it is not limited to manufacturing.

Ross [40] presents work with Hewlett-Packard using AI and IoT to create Digital Twin avatars of people in several ways. From a health perspective, the Digital Twin technology, combined with AI algorithms, can be used to see the effects specific lifestyle changes could have on a person's health, recommending specific changes from AI and Digital Twin analysis. This use emphasises the full integration of data from both the physical twin (The Human) and the Digital Twin (The Replica). Giving the human the ability to see what impact their actions are having on the physical twin while also showing the effect some lifestyle changes could have on them [40].

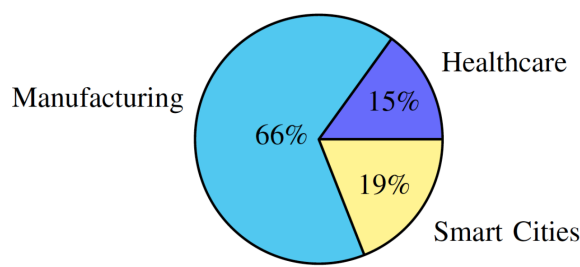

**Figure 5.** Percentage share of research areas found in the analysis of papers presented in Table 5.

In another more recent setting, Laaki et al. [32] present a working prototype of autonomous surgery harnessing IoT and Industry 4.0 connectivity to create a Digital Twin of a patient. The authors propose a remote surgery application through a mobile network. The prototype uses a robotic arm, virtual reality (VR) with a 4G environment, to deliver precision surgery. The paper presents the complexity with multidisciplinary research, citing this as one of the reasons why the work used simulation as opposed to a physical prototype. Laaki et al. also discuss problems with integrating the prototype with a Digital Twin. The authors explore some of the advancements in AI and Industry 4.0 and how they ease the challenges of connectivity, integration and multidisciplinary research [32].

Liu et al. [37] present a novel approach for the future delivery of healthcare combining cloud technology with Digital Twins to create a framework that helps to monitor, diagnose and predict the health of a patient. Liu et al. achieve this through the advancement and use of IoTs wearable technology and in-home sensors with an emphasis on use for the elderly. In addition to a framework, the authors also present several applications citing the feasibility of each. The main contribution of Liu et al.'s paper is the ability to predict a problem with patients more accurately, through the combined use of IoT, Cloud and Digital Twin technology [37].

### 2) Smart Cities

This section is focused on current research involving smart cities in relation to Digital Twins. Research in recent years has seen substantial growth in urbanisation combined with the rise of IoT and data analytics [3]. Mohammadi et al. [3] cite this as one of the motivations for their work and identify the varying states in spatiotemporal flux, emphasising that these need to be understood to maintain growth. The concept they present does this through the use of Digital Twin and virtual reality headsets, allowing them to monitor fluctuations while making predictions through real-time analytics [3].

Table 5. Categorical Review.

| Paper | Type | Defined Twin | Actual Twin | Broad Area | Specific Area | Technology |
|---|---|---|---|---|---|---|
| Bilberg, Malik (2019) [1] | Case Study | DT | DS | Manufacturing | Smart Factory | Simulation |
| Chhetri et al. (2019) [74] | Case Study | DT | DT | Manufacturing | Assembly Line | AI, Sensors, Simulation |
| He et al. (2018) [75] | Review | DT | DS | Manufacturing | Power System | Simulation, AI, Analytics |
| Howard (2019) [11] | Concept | DT | DM | Manufacturing | Product Development | EDA, Visualisation |
| Jain et al. (2019) [76] | Concept | DT | DT | Manufacturing | Fault Diagnosis | Industry 4.0 |
| Karadeniz et al. (2019) [77] | Case Study | DT | DS | Manufacturing | Ice Cream Machines | AR, VR, Industry 4.0, AI, CPS |
| Kuehn (2019) [78] | Concept | DT | DS | Manufacturing | Smart Factory | Simulation |
| Lu (2019) [79] | Review | DT | No Example | Manufacturing | Smart Factory | Cloud, CPS, Industry 4.0 |
| Mandolla et al. (2019) [2] | Case Study | DT | No Example | Manufacturing | Aircraft | Blockchain, Visualisation |
| Mawson, Hughes (2019) [27] | Case Study | DT | DT | Manufacturing | Energy Modelling | Industry 4.0 |
| Min et al. (2019) [35] | Case Study | DT | DS | Manufacturing | Petrochemical Factory | AI, Optimisation |
| Qi, Tao (2018) [24] | Review | DT | DT | Manufacturing | Smart Factory | Industry 4.0, AI, Cloud, Big Data |
| Shangguan et al. (2019) [80] | Case Study | DT | DM | Manufacturing | Wind Turbine | CPS |
| Sivalingam et al. (2018) [17] | Review | DT | DS | Manufacturing | Wind Turbine | CPS, Simulation |
| Tao et al. (2019) [81] | Review | DT | DT | Manufacturing | Smart Factory | CPS, Industry 4.0, AI |
| Tao et al. (2018) [82] | Review | DT | DT | Manufacturing | Assembly Line | CPS, Industry 4.0, AI |
| Xu et al. (2018) [23] | Concept | DT | DS | Manufacturing | Fault Diagnosis | CPS, Industry 4.0, AI, Transfer Learning |
| El Saddik (2018) [39] | Definition | DT | DT | Healthcare | Patient Monitoring | VR, AI |
| Laaki et al. (2019) [32] | Concept | Undefined | DS | Healthcare | Surgery Robotics | Industry 4.0, AI, VR |
| Liu et al (2019) [37] | Concept | DT | DT | Healthcare | Health Management, Elderly Health | Cloud, CPS |
| Ross (2016) [40] | Review | DT | DT | Healthcare | Predictive Health & Well-being | VR, 3D Modelling |
| Chen et al. (2018) [22] | Review | Undefined | DS | Smart City | Driving | Simulation, AI |
| Jo (2018) [15] | Review | DT | DT | Smart City | Livestock Farms | Industry 4.0 |
| Mohammadi, Taylor (2017) [3] | Concept | DT | DT | Smart City | Infrastructure Analysis | Simulation, VR |
| Pargmann et al. (2018) [18] | Review | DT | DS | Smart City | Wind Farm | AR, AI, Cloud |
| Ruohomäki et al. (2018) [83] | Case Study | DT | DS | Smart City | 3D Energy Mapping | Visualisation, Sensors Ontology |

Ruohomäki et al. [83] also present a framework, "myS-MARTlife", which makes use of advancements in IoT across cities to create a smart city Digital Twin. The paper presents a case study for helping in urban planning and built environments, but has particular uses in the energy consumption field with the ability to use the Digital Twin for monitoring and comparing of energy consumption based on the environment and human impact. Both are used for real-time and future developments [83]. Both Mohammadi et al. [3] and Ruohomäki et al. cite the need for the uptake of Industry 4.0 concepts to ensure the level of data exchange is high enough for the twin to perform accurately.

Fuelling the energy grid, along with implementing the integration of renewable energy methods, is a challenge. With this comes the need for accurate delivery within a smart city. Wind power is an example of a renewable energy source which needs to be delivered, monitored and analysed. Pargmann et al. [18] present a cloud-based Digital Twin monitoring system used for the development and monitoring of wind farms. The authors present a working prototype that uses data feeds, and parameters set out both from a technical and business context, allowing for the creation of a working twin of a wind farm development [18].

Sivalingam et al. [17] review and produce a case study that investigates wind farm use and energy consumption for a smart grid. The paper cites some of the challenges with the reliability of power consumption and the general maintenance of wind turbines. The authors propose a working methodology that makes use of IoT sensors combined with data analytics within a Digital Twin environment to accurately perform and predict maintenance of the wind turbines

[17].

In terms of a smart city, Jo et al. [15] and Chen et al. [22] both present work that is related to smart cities and both utilise Digital Twins. The first, by Jo et al. [15], presents a paper that produces a feasibility study on the potential uses of a Digital Twin for a smart farm. The authors note the importance of Industry 4.0 for the realisation of this project as it makes the deployment of Digital Twin a challenge in complex environments. The authors highlight three applications, GE's Predix, Eclipse's Ditto and IBM's Watson as contenders for the deployment of Digital Twin technology along with some guidelines on how the monitoring of livestock can be smarter through the use of a Digital Twin [15].

Chen et al. [22] present Digital Twins for cars and traffic management. The paper explores the challenges with driving, showing there needs to be more data flow within the vehicle used and a connection to other vehicles in the vicinity. Chen et al. [22] present a framework that uses a Digital Twin combined with learning algorithms that monitor and analyse feedback based on user behaviour. The algorithms facilitate a real-time digital behavioural twin of a driver, providing warnings and instructions on how to drive more safely to minimise risk [22]. Both Jo et al. and Chen et al. share the need and challenge of data exchange both citing the need for greater connectivity, which is needed to achieve the optimum result and accuracy for each of the Digital Twins.

3) Manufacturing

Finally, this section concerns an area that includes the majority of research relating to digital twins. It is also an area that has many sub-fields, from large smart factories to smaller machines and tools. For this reason, the section is split into the following subsections: a) Smart Manufacturing Reviews, b) Simulation and Artificial Intelligence, c) System Design and Development and d) Energy Efficient Manufacturing.

a) Smart Manufacturing Reviews

Reviews in this field are limited, but the first to note is presented by Qi and Tao [24] which give a comprehensive "360" view on Digital Twins for big data in a manufacturing and industrial setting. It gives a comparison of enabling technologies for Digital Twins as well as arguing the importance of emerging technologies for the development of smart manufacturing [24].

Tao presents two review papers, [82] and [81]; the first [82] compares Digital Twins and Cyber-Physical systems within a smart manufacturing environment. The latter by Tao [81], is a state-of-the-art paper for Digital Twins combined with industry. Both papers share enabling technologies, IoT, cloud, big data, and artificial intelligence showing how these technologies use Digital Twins. Tao [85] also elaborates on some of the main application uses; Digital Twin simulation, Digital Twin "as a service", data fusion and interaction and collaboration. Both cite the increased development of Industry 4.0 technologies with an emphasis on data analytics and IoT as a factor in the growth and use of Digital Twins. [82] [81] [85].

b) Simulation and Artificial Intelligence

The above papers have touched on some key areas relating to using Digital Twins combined with simulation and AI approaches for manufacturing. Kuehn [78] describes a concept that cites "virtual clones" of a system combined with machine learning algorithms to enhance the manufacturing process. The author categorises key areas in a manufacturing process to highlight their specific goals and concepts for applying a Digital Twin to the manufacturing process, giving enterprises the ability to test, simulate and optimise manufacturing processes in a virtual environment ensuring increased quality and efficiency.

Similarly to Kuehn [78], Min et al. [35] present a paper that exhaustively lists the key enabling technologies with an emphasis on digital twin solutions using AI, specifically machine learning, giving comprehensive evaluations. The work uses a case study to evaluate the pros and cons of using machine learning and Digital Twins for the petrochemical industry, applying each to an industrial IoT petrochemical factory [35]. A limitation found when performing the case study is that the Digital Twin and algorithms are unique to a petrochemical plant. Transfer learning could be used to find commonalities in algorithms to help create solutions transferable to other manufacturing processes.

A subsection of manufacturing is fault diagnosis with two papers of interest found. The first paper by Jain et al. [76] present a simulation study which discusses a Digital Twin approach to fault diagnosis for distributed photovoltaic systems (PV). The advancements in Digital Twin technology allows the team to develop a Digital Twin that can estimate accurately faults relating to PV energy units in real-time [76].

Similar work by Xu et al. [23] proposes a AI focused solution for fault diagnosis in a smart manufacturing environment. Xu et al. highlight challenges in the amount of training data available for creating accurate AI algorithms for new manufacturing processes, something also needed to create accurate Digital Twins. To ensure any challenges are mitigated, the front running of the model is using a Digital Twin which learns and diagnoses faults while producing training data. The second phase can then make use of transfer learning using the collected training data from phase one of the algorithm. A more accurate fault diagnosing system is achieved with the help of a Digital Twin and transfer learning. The authors present a workable case study with a car manufacturing environment testing and evaluating the effectiveness of the concept [23].

c) System Design and Development

Another key area for embracing Digital Twin use is through the design and development stage of manufacturing processes and systems. Shangguan et al. [80] discuss an approach that draws on Digital Twin use but also introduces the Cyber-Physical System (CPS) trend. The authors present a Hierarchical Digital Twin Framework (HDTM) for the design and development of dynamic CPSs in a smart manufacturing environment. The Digital Twin is used throughout the levels of CPS design, harnessing the ability to reuse and test against physical data on a virtual twin. The authors perform a case study showing the benefits of a Digital Twin, which are best seen here as it gives the authors the ability to use the twin in a real-time predictive design setting as well as learning for large scale system changes. The framework is an application for industrial robot design [80].

Howard [11] presents a paper that is an insight into the trend of Digital Twin, evaluating its uses and suitability for the ever-evolving smart manufacturing world. The main goal achieved is the design and development of manufacturing electronic hardware through "virtual validation" utilising a Digital Twin [11]. Both papers cite challenges identified by leaders and visionaries in the world of Digital Twins, analytics and smart manufacturing.

Aside from high-level system design, Karadeniz et al. [77] and Chhetri et al. [74] produce contextual papers and case studies on the development of Digital Twin systems for manufacturing processes. The former [77] discusses the advancements in Industry 4.0 and IoT using a case study to explore how this growth is facilitating the trends of AR (Augmented Reality) and VR (Virtual Reality). The authors set out a concept that creates a Digital Twin of gastronomic things (devices and processes relating to food and cooking) and presents this as "eGastronomic things", similar to the Internet of Things. The gastronomical processes have physical IoT sensors collecting data to create a Digital Twin. The authors use an ice cream machine as a case study to show how a Digital Twin can help in monitoring and maintaining the performance of "eGastronomic" processes, namely the ice cream machine [77].

The latter, by Chhetri et al. [74] also cites the advancements in digitalisation for the growth in smart manufacturing design and developments. The authors propose a methodology and perform a case study that takes advantage of the growth in IoT to build a Digital Twin of a manufacturing process. In this methodology, the IoT sensors help collect and store data streams that are used to indirectly highlight side channel states; acoustics, power and magnetic output of a process. These can be used to localise a fault and identify problems with the manufacturing processes. The team validates the work with a case study of a Fused-Deposition Modelling system (FDM) which produces high accuracy anomaly detection, the first of its kind in this format [74].

Mandolla et al. [2] discuss another trend adapted for the smart manufacturing industry, Blockchain. A growing list of blocks within a decentralised ledger can be used to record data across many computers. The previous blocks are linked using a cryptography hash, a method of protecting data using encrypted codes. These hashes are unique to each block and contain attributes such as timestamp and transaction data. Blockchain growth is attributed to its increased use in processes other than its original intended use; for example, cryptocurrency [86] [87] [88]. Mandolla et al. [2] present an example of blockchain use, in a case study showing how they have combined it with a Digital Twin. The authors focus on aerospace manufacturing concerning the metal additive process. Mandolla et al. create a Digital Twin of this process while providing a conceptual answer to securing the process through blockchain, and monitoring through a Digital Twin [2]. Similarly, Bilberg and Malik [1] are using robotics, to explore how they can be adapted for a manufacturing process, they present an event driven Digital Twin that works in parallel with a robot to perform tasks on an assembly line. The other combines the physical task and the virtual task to create a real-time, skills-based robotic production line accurately allocating a task to the human or robot based on the optimum production they could respectively achieve [1]. Both papers, through the use of case studies, identify the benefits of digitalisation, but both also cite the challenges with seamless integration, something needed for the effective running of an assembly line and Digital Twins [2] [1] [89].

d) Energy Efficient Manufacturing

In line with national and international targets, energy used for manufacturing, is required to be monitored and reduced; hence, the need for potential solutions for energy efficient production lines. Both Lu et al. [79] and Mawson [27] present systems and architecture for energy efficient manufacturing. A driver for this are the environmental benefits. However, more efficient manufacturing will also reduce cost, increase profits and future investments. Lu et al. present a paper that focuses on developing an architecture that implements energy aware Digital Twin model with a platform called MCLoud. Both of which facilitate an Industry 4.0 environment where manufacturing processes are monitored continuously and self-configured through CPSs and Digital Twins, with energy efficiency the overall goal.

Mawson et al. [27] also draw on the advancement in Industry 4.0, citing a review on their effectiveness for increased automation, connectivity and flexibility for manufacturing processes. The main contributions of the paper are a holistic review and case study using methodologies and frameworks derived from the analysis of energy consumption at the machine process level. Mawson et al. cite various simulation tools, but they lack multi-level integration.

To produce a high accuracy model, all aspects of the manufacturing process from the materials to resource flows are needed to analyse energy consumption accurately. These are challenges demonstrating how Digital Twins, VR and AR will facilitate future research [27].

## VI. OPEN RESEARCH

The penultimate section in this review briefly discusses open research questions for Digital Twins as well as exsploring the literature reviewed for challenges facing future research.

### A. DIGITAL TWIN IN MANUFACTURING

Manufacturing and fully integrated Digital Twins are proposed in the literature, but are not currently realised in industry. Section V, above, explored a number of case studies, reviews and concepts relating to publications in the manufacturing industry. There are a number of publications on smaller developments of a Digital Twin. Industry examples of case studies were listed in Table 5 and discussed in greater detail in Section V subsection C and summarised below.

In the literature not all publications cover all aspects of a Digital Twin from the physical and virtual modelling to the data, connection and service parts of modelling a Digital Twin. Modelling and scaling are needed to create generic Digital Twins. Zheng et al. [8] and Schleich et al. [34] both present sensor-based modelling of different stages of a Digital Twin with Shangguam et al. [80] only focusing on the hierarchy modelling of the virtual and physical modelling of a Digital Twin; there is a need for a generic model for a complete Digital Twin.

Literature covering data fusion is another topic that is heavily researched across all areas of science but is less researched when applying data fusion and Digital Twins in an industrial setting. Likewise, with Digital Twin modelling, another avenue is to incorporate data fusion when developing generic models. The literature explores how predictive maintenance can incorporate Digital Twins with data fusion which is a promising area of research [90] [7].

Lacking in research is the modelling of Digital Twins that use virtual and physical data fusion, this can also be discussed under the term "Cyber-Physical" fusion. The need is for standardised approaches when modelling data fusion with Digital Twins. [90] [7] and [91], explore some of the ways data fusion can be used for Digital Twins in industry, while also highlighting some of the potential challenges with implementing data fusion, from connection problems to security threats.

Years of research in the maintenance of machines has produced the term prognostics and health management (PHM) primarily used in an industrial setting as it can be applied to the health and manufacturing processes from small to large scale plants [92] [93] [23]. Digital Twin technology promotes PHM as a potential for research in areas of fault diagnosis and predictive maintenance for industrial processes, which are tangible with the development of Digital Twins [10] [94].

The Digital Twin also allows for the potential interaction and collaboration of machines, giving the ability for simulation of processes, facilitating the goal of more accurate manufacturing [95] [96] [82].

#### 1) Industry Case Studies

In the literature reviewed above, seven authors present findings from a number of case studies on Digital Twin implementations. To summarise [1] [35] and [2], all three papers and case studies discuss how digital twins can be utilised to optimise the manufacturing processes. [1] explores the benefits in optimisation when applying a Digital Twin to an assembly line. Min et al. [35] investigate how Digital Twin use can optimise production in the petrochemical industry. [2] presents work on how Digital Twins and blockchain can be used to support and optimise the aircraft manufacturing industry.

The first three publications above concern case studies which explore optimisation, while the next four, [74] [77] [27] and [80], relate to the modelling of Digital Twin systems. [74] investigates a case study on how to model and maintain a Digital Twin with IoT sensors; [27] uses a case study to explore how to develop digital twin models using a multi-levelled framework and [80] explores the use of a hierarchical modelling framework for Digital Twin development. The last paper, by Karandiz et al. [77] presents a case study on the 3D modelling of an ice cream machine exploring how a Digital Twin can be utilised.

Specific findings from each of these case studies contribute to the overall finding from the literature, presented in Section VI.D below.

### B. DIGITAL TWIN IN HEALTHCARE

This next section discusses the open research questions for Digital Twins in a healthcare setting. Some of the research cites the potential for adapting Digital Twin technology for humans. An example is a Digital Twin of a person to monitor day-to-day health and well-being giving the potential for a human twin for simulating what positive and negative lifestyle changes could have on the physical human. The significant open research comes in the form of modelling and breaking down the barriers to modelling a human body to a Digital Twin. Again, down to the issue of having no standardised Digital Twin modelling methods [91]. Similar to PHM in manufacturing, this concept is an exciting area of research with the Digital Twin being used to monitor and maintain people's health. From day-to-day healthcare to ongoing health conditions, the Digital Twin can be used in a similar way to PHM, combining data analytics to ensure patients are healthy. [32] presents open research relating to surgery; using historical data with current real-time data for Digital Twin simulations of surgery and overall healthcare. The aim is to spot risks before they arise using the virtual twin of the patient.

An area of research is in the field of data fusion, with [32] and [37] citing the need for research in accurately dealing with data collected and processed for a Digital Twin, mainly as it deals with sensitive patient data over the virtual and physical Digital Twin, adequate interaction and convergence are needed for greater trust and use; hence, the need for more research.

Remote surgery and healthcare is another exciting area of research. The ability of a doctor being able to perform pre-surgery checks remotely through a Digital Twin is a promising way to minimise risk to life. Another concept presented by Lakki et al. [32] is the open research for network supported remote surgery; this comes in line with the developments of 5G for mobile networked surgery, another area of future research for Digital Twins in healthcare [32] [40].

Data fusion, modelling, remote surgery and the implementation of Digital Twins for healthcare facilitates more specific areas of research. [40] [39] [32] and [37] all cite the ongoing concern with the security of collection and processing of data for a Digital Twins, more importantly when dealing with sensitive data from a healthcare setting. The goal of future work is to ensure the privacy of the data used for a Digital Twin.

### C. DIGITAL TWIN IN SMART CITIES

The final area of interest is the open research questions for Digital Twins assosiated with smart cities. The review reported in Section V, above, shows that this area is similar to healthcare in terms of limited academic research. The papers currently cover large parts of this broader vision of a smart city; anything from city-wide Digital Twins for smart cities to more specific areas like traffic management Digital Twins [22] [97] and livestock management [15] to renewable energy [18]. The topics touched on how the large scale Digital Twin can be used for a fully connected large town or city.

An exciting opportunity is the combination of Digital Twin and local infrastructure, as explored by [3] and [83], making use of 3D modelling for smart city development and maintenance. Open areas to be researched come in the form of applying data analytics, such as predictive analytics applied to a Digital Twin for developing smart city.

A standardised Digital Twin for a smart city is a must. However, this necessitates further research in areas for the interactions of data fusion for the physical and virtual interchange of data for a Digital Twin. The need for modelling in a smart city is emphasised by [22] and [97] with the call for more generic models that take into account all components of a smart city. Chen [22] presents a traffic management concept integrated without the need for further developments or more generic modelling, setting a standard for compatibility between smart buildings and smart traffic. With the rise and development of Digital Twins for manufacturing, it is clear to see the opportunities for smart cities are increasing [3] [83] [22].

### D. COMMON NEEDS AND SPECIFIC FINDINGS

#### 1) Data Models

The first set of findings relate to the Digital Twin model and its architecture. There is a lack of unified models or a generic Digital Twin architecture in the literature, with no consensus on how to build a Digital Twin system. Developing a design paradigm for building a Digital Twin could be a generic way of implementing a basic Digital Twin system.

#### 2) Heterogeneous Systems

As the Digital Twin environment is heterogeneous, while also being connected to large distributed networks, it is an ongoing goal seen throughout the field for researchers to gain a deeper understanding of how to deal with such systems. This is something to be achieved through evaluating and comparing current Digital Twin systems with each other, exploring how they differ in handling smaller and larger heterogeneous Digital Twin environments.

#### 3) Artificial Intelligence (AI)

With the advancement in digitalisation, AI is one of the leaders and facilitators for growth and adaptability for enabling Digital Twins and is becoming the main component of such systems. The open areas of research evaluating the impact of AI, machine and deep learning algorithms could potentially have on the advancements and uses of Digital Twin technology. These advancements in Industry 4.0 concepts have enriched the way we live, work and communicate, in turn, opening up more research opportunities as the demand for Digital Twin technology increases. AI paves the way for more case studies in an industry setting for evaluating the predictive maintenance and remaining useful life of systems as a case study.

#### 4) Security

With the advancements in technology like blockchain, come several opportunities to ensure security with the main focus on exploring solutions that aid secure Digital Twins.

#### 5) Data Exchange

Like security and AI, with more connectivity comes the rise in Digital Twin systems and its enabling technologies. More research will ensure the Digital Twin can thrive when sharing data between devices. An open area is looking for solutions in achieving the seamless integration of data for small IoT systems as well as large heterogeneous systems. The development of small Digital Twins needs to be scaled up, with consideration for weak data exchange.

#### 6) IoT

Most of the open research is from an IoT perspective when thinking of a Digital Twins. There needs to be a way to retrofit sensors to ensure the data exchange is accurate and performing to its best ability. Edge computing is another open area of research for IoT and Digital Twin technology.

### E. CHALLENGES TO OPEN RESEARCH

There are a number of challenges encountered when trying to address open research questions for Digital Twins, as discussed below.

#### 1) Multidisciplinarity

One significant challenge comes from the multidisciplinary environments used for designing and developing. These arise due to the many different fields of research involved in collaborating, which can aid breakthroughs, but also a hindrance as goals in several fields lead in different directions for the research, ultimately leading to slower results.

### 2) Standardisation

As highlighted in this paper and seen in many new and emerging technologies, another challenge facing Digital Twin is with the lack of standardisation. This results in discrepancies between projects in Digital Twin. A contributing factor is the variety of definitions seen; this, coupled with no standardisation is a challenge, slowing the progress of the Digital Twin technology.

### 3) Global Advancements

With the vast advancements in Digital Twins and their enabling technologies, come many benefits but also challenges. Futuristic and unrealistic goals lead to slower uptake and development of newly adapted technologies and opinions. With the breakthrough of Industry 4.0 and the rapid growth of digitalisation comes the growth of technologies at different rates. However, the accelerated growth leads to a range of challenges with connectivity and data exchange. For example, the supporting systems or IoT devices may not be compatible with the Digital Twin. This growth at different rates can also pose new challenges for security in terms of exposing vulnerabilities.

## VII. CONCLUSION

The growth in Digital Twin use has seen a shift in recent years, facilitated by an increase in the number of published papers and industry leaders investing heavily in developing Digital Twin technology. It would not be possible without the same growth in the AI, IoT and IIoT fields, which are becoming key enablers for Digital Twins. The majority of the Digital Twin research is focused on the manufacturing field, as evidenced through the large proportion of papers in this area reviewed above. The number of papers found in manufacturing is noticeably higher compared to papers discussing Digital Twins for smart cities and healthcare, highlighting gaps in the research for these areas.

AI is becoming a component within Digital Twins and exploring where these algorithms can be applied is another avenue of open research. The effects of AI combined with Digital Twin are topics amongst the publications but on a small scale. The exciting and inevitable future research will explore scaling up smaller successful Digital Twin and AI projects. An important finding is the lack of standardisation and misconceptions with definitions for Digital Twins. Addressing the challenges with standardisation ensures future developments are actually Digital Twins and not wrongly defined concepts.

The review carried out above highlights two other areas of growing interest, Digital Twins for healthcare and smart cities. Thus, the reason why the paper contributes to a categorical review that includes not only manufacturing but healthcare and smart cities. The paper discusses each area, highlighting how researchers are developing Digital Twins, while also identifying challenges and key enabling technologies, thus aiding future work. The paper also identifies the lack of clear definitions for a Digital Twin, showing how there is no real difference in definition since being initially coined in 2012. It is also evident that some research wrongly identifies Digital Twins as models and shadows. Across the literature, there are examples of small scale Digital Twin projects, but a lack of large scale projects. One reason for this is the lack of domain knowledge on successfully scaling up larger Digital Twins. Papers concerning Digital Twin use in manufacturing identify a range of publications with particular growth in the health of the machines and predictive maintenance. Digital Twins for healthcare draws on similar themes in terms of health status and monitoring, with a number of papers investigating Digital Twins use for predictive analytics of human users. The paper also highlights the advancements in remote surgery and the importance of researching data fusion, mainly due to the nature of sensitive data used in healthcare. Research for smart cities is limited, but the potential to investigate Digital Twins for traffic management systems and smart city developments is on the rise.

Despite the field of Digital Twin being in its infancy and dominated by manufacturing, this review paves the way for further work. The paper provides a foundation for other researchers to investigate the field further.

## ACKNOWLEDGEMENT

This work is partly supported by the SEND project (grant ref. 32R16P00706) funded by ERDF and BEIS. We would also like to thank Astec IT Solution Ltd as an industry partner in conducting this research.

## References

[1] A. Bilberg and A. A. Malik, "Digital twin driven human–robot collaborative assembly," CIRP Annals, Apr. 2019.
[2] C. Mandolla, A. M. Petruzzelli, G. Percoco, and A. Urbinati, "Building a digital twin for additive manufacturing through the exploitation of blockchain: A case analysis of the aircraft industry," Computers in Industry, vol. 109, pp. 134–152, Aug. 2019.
[3] N. Mohammadi and J. E. Taylor, "Smart city digital twins," in 2017 IEEE Symposium Series on Computational Intelligence (SSCI), (Honolulu, HI), pp. 1–5, IEEE, Nov. 2017.
[4] M. Grieves, "Digital Twin: Manufacturing Excellence through Virtual Factory Replication," white Paper, NASA, 2014.
[5] E. Glaessgen and D. Stargel, "The Digital Twin Paradigm for Future NASA and U.S. Air Force Vehicles," in 53rd AIAA Structures, Structural Dynamics and Materials Conference, (Honolulu, Hawaii), American Institute of Aeronautics and Astronautics, Apr. 2012.
[6] Y. Chen, "Integrated and Intelligent Manufacturing: Perspectives and Enablers," Engineering, vol. 3, pp. 588–595, Oct. 2017.
[7] Z. Liu, N. Meyendorf, and N. Mrad, "The role of data fusion in predictive maintenance using digital twin," in Annual Review Of Progress In Quantitative Nondestructive Evaluation, (Provo, Utah, USA), p. 020023, 2018.
[8] Y. Zheng, S. Yang, and H. Cheng, "An application framework of digital twin and its case study," Journal of Ambient Intelligence and Humanized Computing, vol. 10, pp. 1141–1153, June 2018.
[9] R. Vrabič, J. A. Erkoyuncu, P. Butala, and R. Roy, "Digital twins: Understanding the added value of integrated models for through-life engineering services," Procedia Manufacturing, vol. 16, pp. 139–146, 2018.
[10] A. Madni, C. Madni, and S. Lucero, "Leveraging Digital Twin Technology in Model-Based Systems Engineering," Systems, vol. 7, p. 7, Jan. 2019.
[11] D. Howard, "The Digital Twin: Virtual Validation In Electronics Development And Design," in 2019 Pan Pacific Microelectronics Symposium (Pan Pacific), (Kauai, HI, USA), pp. 1–9, IEEE, Feb. 2019.
[12] A. Coraddu, L. Oneto, F. Baldi, F. Cipollini, M. Atlar, and S. Savio, "Data-driven ship digital twin for estimating the speed loss caused by the marine fouling," Ocean Engineering, vol. 186, p. 106063, Aug. 2019.

[13] J. David, A. Lobov, and M. Lanz, "Leveraging Digital Twins for Assisted Learning of Flexible Manufacturing Systems," in 2018 IEEE 16th International Conference on Industrial Informatics (INDIN), (Porto), pp. 529–535, IEEE, July 2018.

[14] T. DebRoy, W. Zhang, J. Turner, and S. Babu, "Building digital twins of 3d printing machines," Scripta Materialia, vol. 135, pp. 119–124, July 2017.

[15] S.-K. Jo, D.-H. Park, H. Park, and S.-H. Kim, "Smart Livestock Farms Using Digital Twin: Feasibility Study," in 2018 International Conference on Information and Communication Technology Convergence (ICTC), (Jeju), pp. 1461–1463, IEEE, Oct. 2018.

[16] G. Knapp, T. Mukherjee, J. Zuback, H. Wei, T. Palmer, A. De, and T. DebRoy, "Building blocks for a digital twin of additive manufacturing," Acta Materialia, vol. 135, pp. 390–399, Aug. 2017.

[17] K. Sivalingam, M. Sepulveda, M. Spring, and P. Davies, "A Review and Methodology Development for Remaining Useful Life Prediction of Offshore Fixed and Floating Wind turbine Power Converter with Digital Twin Technology Perspective," in 2018 2nd International Conference on Green Energy and Applications (ICGEA), (Singapore), pp. 197–204, IEEE, Mar. 2018.

[18] H. Pargmann, D. Euhausen, and R. Faber, "Intelligent big data processing for wind farm monitoring and analysis based on cloud-technologies and digital twins: A quantitative approach," in 2018 IEEE 3rd International Conference on Cloud Computing and Big Data Analysis (ICCCBDA), (Chengdu), pp. 233–237, IEEE, Apr. 2018.

[19] C. Brosinsky, D. Westermann, and R. Krebs, "Recent and prospective developments in power system control centers: Adapting the digital twin technology for application in power system control centers," in 2018 IEEE International Energy Conference (ENERGYCON), (Limassol), pp. 1–6, IEEE, June 2018.

[20] H. Brandtstaedter, C. Ludwig, L. Hubner, E. Tsouchnika, A. Jungiewicz, and U. Wever, "DIGITAL TWINS FOR LARGE ELECTRIC DRIVE TRAINS," in 2018 Petroleum and Chemical Industry Conference Europe (PCIC Europe), (Antwerp), pp. 1–5, IEEE, June 2018.

[21] R.-M. Soe, "FINEST Twins: platform for cross-border smart city solutions," in Proceedings of the 18th Annual International Conference on Digital Government Research - dg.o '17, (Staten Island, NY, USA), pp. 352–357, ACM Press, 2017.

[22] X. Chen, E. Kang, S. Shiraishi, V. M. Preciado, and Z. Jiang, "Digital Behavioral Twins for Safe Connected Cars," in Proceedings of the 21st ACM/IEEE International Conference on Model Driven Engineering Languages and Systems - MODELS '18, (Copenhagen, Denmark), pp. 144–153, ACM Press, 2018.

[23] Y. Xu, Y. Sun, X. Liu, and Y. Zheng, "A Digital-Twin-Assisted Fault Diagnosis Using Deep Transfer Learning," IEEE Access, vol. 7, pp. 19990–19999, 2019.

[24] Q. Qi and F. Tao, "Digital Twin and Big Data Towards Smart Manufacturing and Industry 4.0: 360 Degree Comparison," IEEE Access, vol. 6, pp. 3585–3593, 2018.

[25] M. Kritzler, J. Hodges, D. Yu, K. Garcia, H. Shukla, and F. Michahelles, "Digital Companion for Industry," in Companion Proceedings of The 2019 World Wide Web Conference on - WWW '19, (San Francisco, USA), pp. 663–667, ACM Press, 2019.

[26] Y. Umeda, J. Ota, F. Kojima, M. Saito, H. Matsuzawa, T. Sukekawa, A. Takeuchi, K. Makida, and S. Shirafuji, "Development of an education program for digital manufacturing system engineers based on 'Digital Triplet' concept," Procedia Manufacturing, vol. 31, pp. 363–369, 2019.

[27] V. J. Mawson and B. R. Hughes, "The development of modelling tools to improve energy efficiency in manufacturing processes and systems," Journal of Manufacturing Systems, vol. 51, pp. 95–105, Apr. 2019.

[28] R. He, G. Chen, C. Dong, S. Sun, and X. Shen, "Data-driven digital twin technology for optimized control in process systems," ISA Transactions, May 2019.

[29] D. Burnett, J. Thorp, D. Richards, K. Gorkovenko, and D. Murray-Rust, "Digital twins as a resource for design research," in Proceedings of the 8th ACM International Symposium on Pervasive Displays - PerDis '19, (Palermo, Italy), pp. 1–2, ACM Press, 2019.

[30] F. Longo, L. Nicoletti, and A. Padovano, "Ubiquitous knowledge empowers the Smart Factory: The impacts of a Service-oriented Digital Twin on enterprises' performance," Annual Reviews in Control, vol. 47, pp. 221–236, 2019.

[31] M. Lohtander, N. Ahonen, M. Lanz, J. Ratava, and J. Kaakkunen, "Micro Manufacturing Unit and the Corresponding 3d-Model for the Digital Twin," Procedia Manufacturing, vol. 25, pp. 55–61, 2018.

[32] H. Laaki, Y. Miche, and K. Tammi, "Prototyping a Digital Twin for Real Time Remote Control Over Mobile Networks: Application of Remote Surgery," IEEE Access, vol. 7, pp. 20325–20336, 2019.

[33] H. Zhang, Q. Liu, X. Chen, D. Zhang, and J. Leng, "A Digital Twin-Based Approach for Designing and Multi-Objective Optimization of Hollow Glass Production Line," IEEE Access, vol. 5, pp. 26901–26911, 2017.

[34] B. Schleich, N. Anwer, L. Mathieu, and S. Wartzack, "Shaping the digital twin for design and production engineering," CIRP Annals, vol. 66, no. 1, pp. 141–144, 2017.

[35] Q. Min, Y. Lu, Z. Liu, C. Su, and B. Wang, "Machine Learning based Digital Twin Framework for Production Optimization in Petrochemical Industry," International Journal of Information Management, May 2019.

[36] M. Joordens and M. Jamshidi, "On The Development of Robot Fish Swarms in Virtual Reality with Digital Twins," in 2018 13th Annual Conference on System of Systems Engineering (SoSE), (Paris), pp. 411–416, IEEE, June 2018.

[37] Y. Liu, L. Zhang, Y. Yang, L. Zhou, L. Ren, F. Wang, R. Liu, Z. Pang, and M. J. Deen, "A Novel Cloud-Based Framework for the Elderly Healthcare Services Using Digital Twin," IEEE Access, vol. 7, pp. 49088–49101, 2019.

[38] Sonal, S. Reddy, and D. Kumar, "Review of Smart Health Monitoring Approaches with Survey Analysis and Proposed Framework," IEEE Internet of Things Journal, pp. 1–1, 2018.

[39] A. El Saddik, "Digital Twins: The Convergence of Multimedia Technologies," IEEE MultiMedia, vol. 25, pp. 87–92, Apr. 2018.

[40] D. Ross, "Digital twinning [Information Technology Virtual Reality]," Engineering Technology, vol. 11, pp. 44–45, May 2016.

[41] R. Magargle, L. Johnson, P. Mandloi, P. Davoudabadi, O. Kesarkar, S. Krishnaswamy, J. Batteh, and A. Pitchaikani, "A Simulation-Based Digital Twin for Model-Driven Health Monitoring and Predictive Maintenance of an Automotive Braking System," in International Modelica Conference, pp. 35–46, July 2017.

[42] D. Petrik and G. Herzwurm, "iIoT ecosystem development through boundary resources: a Siemens MindSphere case study," in Proceedings of the 2nd ACM SIGSOFT International Workshop on Software-Intensive Business: Start-ups, Platforms, and Ecosystems - IWSiB 2019, (Tallinn, Estonia), pp. 1–6, ACM Press, 2019.

[43] S. Kumar and A. Jasuja, "Air quality monitoring system based on IoT using Raspberry Pi," in 2017 International Conference on Computing, Communication and Automation (ICCCA), pp. 1341–1346, May 2017.

[44] V. Damjanovic-Behrendt, "A Digital Twin-based Privacy Enhancement Mechanism for the Automotive Industry," in 2018 International Conference on Intelligent Systems (IS), pp. 272–279, Sept. 2018.

[45] Automation, "Bentley Systems releases iModel.js open-source library," Oct. 2018.

[46] R. Vishwakarma and A. K. Jain, "A survey of DDoS attacking techniques and defence mechanisms in the IoT network," Telecommunication Systems, July 2019.

[47] K. Ashton, "That 'Internet of Things' Thing - 2009-06-22 - Page 1 - RFID Journal," RFID Journal, June 2009.

[48] S. Ornes, "Core Concept: The Internet of Things and the explosion of interconnectivity," Proceedings of the National Academy of Sciences, vol. 113, pp. 11059–11060, Oct. 2016.

[49] Knud Lasse Lueth, "State of the IoT 2018: Number of IoT devices now at 7b – Market accelerating."

[50] Statista, "IoT: number of connected devices worldwide 2012-2025."

[51] David Curry, "ARM: One trillion IoT devices by 2035, $5 trillion in market value," July 2017.

[52] I. U. Din, M. Guizani, J. J. Rodrigues, S. Hassan, and V. V. Korotaev, "Machine learning in the Internet of Things: Designed techniques for smart cities," Future Generation Computer Systems, vol. 100, pp. 826–843, Nov. 2019.

[53] H. Boyes, B. Hallaq, J. Cunningham, and T. Watson, "The industrial internet of things (IIoT): An analysis framework," Computers in Industry, vol. 101, pp. 1–12, Oct. 2018. *** goooood good.

[54] K. Kobara, "Cyber Physical Security for Industrial Control Systems and IoT," IEICE Transactions on Information and Systems, vol. E99.D, no. 4, pp. 787–795, 2016.

[55] X. Fei, N. Shah, N. Verba, K.-M. Chao, V. Sanchez-Anguix, J. Lewandowski, A. James, and Z. Usman, "CPS data streams analytics based on machine learning for Cloud and Fog Computing: A survey," Future Generation Computer Systems, vol. 90, pp. 435–450, Jan. 2019.

[56] Y. Liao, E. d. F. R. Loures, and F. Deschamps, "Industrial Internet of Things: A Systematic Literature Review and Insights," IEEE Internet of Things Journal, vol. 5, pp. 4515–4525, Dec. 2018.

[57] S V Engineering College, Tirupati, India and R. R. B, "A Comprehensive Literature Review on Data Analytics in IIoT (Industrial Internet of Things)," HELIX, vol. 8, pp. 2757–2764, Jan. 2018.

[58] M. Stanley, "IIoT & the New Industrial Revolution," 2016.

[59] "Digital Twin as Enabler for an Innovative Digital Shopfloor Management System in the ESB Logistics Learning Factory at Reutlingen - University | Elsevier Enhanced Reader." Resersch question.

[60] NetObjex, "Industrial Revolution," 2018.

[61] A. Čolaković and M. Hadžialić, "Internet of Things (IoT): A review of enabling technologies, challenges, and open research issues," Computer Networks, vol. 144, pp. 17–39, Oct. 2018.

[62] M. M. Najafabadi, F. Villanustre, T. M. Khoshgoftaar, N. Seliya, R. Wald, and E. Muharemagic, "Deep learning applications and challenges in big data analytics," Journal of Big Data, vol. 2, Dec. 2015.

[63] I. Taleb, R. Dssouli, and M. A. Serhani, "Big Data Pre-processing: A Quality Framework," in 2015 IEEE International Congress on Big Data, pp. 191–198, June 2015. ISSN: 2379-7703.

[64] Q. Qi, F. Tao, Y. Zuo, and D. Zhao, "Digital Twin Service towards Smart Manufacturing," Procedia CIRP, vol. 72, pp. 237–242, 2018.

[65] S. Srivastava, A. Bisht, and N. Narayan, "Safety and security in smart cities using artificial intelligence — A review," in 2017 7th International Conference on Cloud Computing, Data Science Engineering - Confluence, pp. 130–133, Jan. 2017.

[66] O. Elijah, T. A. Rahman, I. Orikumhi, C. Y. Leow, and M. N. Hindia, "An Overview of Internet of Things (IoT) and Data Analytics in Agriculture: Benefits and Challenges," IEEE Internet of Things Journal, vol. 5, pp. 3758–3773, Oct. 2018.

[67] O. B. Sezer, E. Dogdu, and A. M. Ozbayoglu, "Context-Aware Computing, Learning, and Big Data in Internet of Things: A Survey," IEEE Internet of Things Journal, vol. 5, pp. 1–27, Feb. 2018.

[68] A. Moujahid, M. E. Tantaoui, M. D. Hina, A. Soukane, A. Ortalda, A. ElKhadimi, and A. Ramdane-Cherif, "Machine Learning Techniques in ADAS: A Review," in 2018 International Conference on Advances in Computing and Communication Engineering (ICACCE), pp. 235–242, June 2018.

[69] P. P. Shinde and S. Shah, "A Review of Machine Learning and Deep Learning Applications," in 2018 Fourth International Conference on Computing Communication Control and Automation (ICCUBEA), pp. 1–6, Aug. 2018.

[70] A. S. Modi, "Review Article on Deep Learning Approaches," in 2018 Second International Conference on Intelligent Computing and Control Systems (ICICCS), pp. 1635–1639, June 2018.

[71] K. E. Barkwell, A. Cuzzocrea, C. K. Leung, A. A. Ocran, J. M. Sanderson, J. A. Stewart, and B. H. Wodi, "Big Data Visualisation and Visual Analytics for Music Data Mining," in 2018 22nd International Conference Information Visualisation (IV), pp. 235–240, July 2018.

[72] S. E. Bibri and J. Krogstie, "The core enabling technologies of big data analytics and context-aware computing for smart sustainable cities: a review and synthesis," Journal of Big Data, vol. 4, Dec. 2017.

[73] W. Kritzinger, M. Karner, G. Traar, J. Henjes, and W. Sihn, "Digital Twin in manufacturing: A categorical literature review and classification," IFAC-PapersOnLine, vol. 51, no. 11, pp. 1016–1022, 2018.

[74] S. R. Chhetri, S. Faezi, A. Canedo, and M. A. A. Faruque, "QUILT: quality inference from living digital twins in IoT-enabled manufacturing systems," in Proceedings of the International Conference on Internet of Things Design and Implementation - IoTDI '19, (Montreal, Quebec, Canada), pp. 237–248, ACM Press, 2019.

[75] Y. He, J. Guo, and X. Zheng, "From Surveillance to Digital Twin: Challenges and Recent Advances of Signal Processing for Industrial Internet of Things," IEEE Signal Processing Magazine, vol. 35, pp. 120–129, Sept. 2018.

[76] P. Jain, J. Poon, J. P. Singh, C. Spanos, S. Sanders, and S. K. Panda, "A Digital Twin Approach for Fault Diagnosis in Distributed Photovoltaic System," IEEE Transactions on Power Electronics, pp. 1–1, 2019.

[77] A. M. Karadeniz, I. Arif, A. Kanak, and S. Ergun, "Digital Twin of eGastronomic Things: A Case Study for Ice Cream Machines," in 2019 IEEE International Symposium on Circuits and Systems (ISCAS), (Sapporo, Japan), pp. 1–4, IEEE, May 2019.

[78] W. Kuehn, "Simulation in Digital Enterprises," in Proceedings of the 11th International Conference on Computer Modeling and Simulation - ICCMS 2019, (North Rockhampton, QLD, Australia), pp. 55–59, ACM Press, 2019.

[79] Y. Lu, T. Peng, and X. Xu, "Energy-efficient cyber-physical production network: Architecture and technologies," Computers & Industrial Engineering, vol. 129, pp. 56–66, Mar. 2019.

[80] D. Shangguan, L. Chen, and J. Ding, "A Hierarchical Digital Twin Model Framework for Dynamic Cyber-Physical System Design," in Proceedings of the 5th International Conference on Mechatronics and Robotics Engineering - ICMRE'19, (Rome, Italy), pp. 123–129, ACM Press, 2019.

[81] F. Tao, Q. Qi, L. Wang, and A. Nee, "Digital Twins and Cyber–Physical Systems toward Smart Manufacturing and Industry 4.0: Correlation and Comparison," Engineering, vol. 5, pp. 653–661, Aug. 2019.

[82] F. Tao, J. Cheng, Q. Qi, M. Zhang, H. Zhang, and F. Sui, "Digital twin-driven product design, manufacturing and service with big data," The International Journal of Advanced Manufacturing Technology, vol. 94, pp. 3563–3576, Feb. 2018.

[83] T. Ruohomäki, E. Airaksinen, P. Huuska, O. Kesäniemi, M. Martikka, and J. Suomisto, "Smart City Platform Enabling Digital Twin," in 2018 International Conference on Intelligent Systems (IS), pp. 155–161, Sept. 2018.

[84] J. Liu, X. Du, H. Zhou, X. Liu, L. ei Li, and F. Feng, "A digital twin-based approach for dynamic clamping and positioning of the flexible tooling system," Procedia CIRP, vol. 80, pp. 746–749, 2019.

[85] F. Tao, H. Zhang, A. Liu, and A. Y. C. Nee, "Digital Twin in Industry: State-of-the-Art," IEEE Transactions on Industrial Informatics, vol. 15, pp. 2405–2415, Apr. 2019.

[86] S. Huh, S. Cho, and S. Kim, "Managing IoT devices using blockchain platform," in 2017 19th International Conference on Advanced Communication Technology (ICACT), pp. 464–467, Feb. 2017.

[87] O. Novo, "Blockchain Meets IoT: An Architecture for Scalable Access Management in IoT," IEEE Internet of Things Journal, vol. 5, pp. 1184–1195, Apr. 2018.

[88] A. Reyna, C. Martín, J. Chen, E. Soler, and M. Díaz, "On blockchain and its integration with IoT. Challenges and opportunities," Future Generation Computer Systems, vol. 88, pp. 173–190, Nov. 2018.

[89] M. T. Malik, Ali Ahmad and A. Bilberg, "Virtual reality in manufacturing: immersive and collaborative artificial-reality in design of human-robot workspace," International Journal of Computer Integrated Manufacturing, Jan. 2020.

[90] Y. Cai, B. Starly, P. Cohen, and Y.-S. Lee, "Sensor Data and Information Fusion to Construct Digital-twins Virtual Machine Tools for Cyber-physical Manufacturing," Procedia Manufacturing, vol. 10, pp. 1031–1042, 2017.

[91] F. Tao and M. Zhang, "Digital Twin Shop-Floor: A New Shop-Floor Paradigm Towards Smart Manufacturing," IEEE Access, vol. 5, pp. 20418–20427, 2017.

[92] E. Negri, L. Fumagalli, and M. Macchi, "A Review of the Roles of Digital Twin in CPS-based Production Systems," Procedia Manufacturing, vol. 11, pp. 939–948, 2017.

[93] K. M. Alam and A. El Saddik, "C2ps: A Digital Twin Architecture Reference Model for the Cloud-Based Cyber-Physical Systems," IEEE Access, vol. 5, pp. 2050–2062, 2017.

[94] R. Stark, C. Fresemann, and K. Lindow, "Development and operation of Digital Twins for technical systems and services," CIRP Annals, May 2019.

[95] D. Wu, C. Jennings, J. Terpenny, and S. Kumara, "Cloud-based machine learning for predictive analytics: Tool wear prediction in milling," in 2016 IEEE International Conference on Big Data (Big Data), pp. 2062–2069, Dec. 2016.

[96] F. Tao, J. Cheng, and Q. Qi, "IIHub: An Industrial Internet-of-Things Hub Toward Smart Manufacturing Based on Cyber-Physical System," IEEE Transactions on Industrial Informatics, vol. 14, pp. 2271–2280, May 2018.

[97] R.-M. Soe, "Smart Twin Cities via Urban Operating System," in Proceedings of the 10th International Conference on Theory and Practice of Electronic Governance - ICEGOV '17, (New Delhi AA, India), pp. 391–400, ACM Press, 2017.

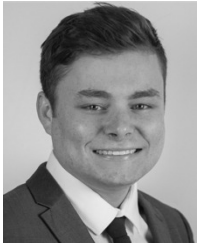

AIDAN FULLER received a Bachelor of Science degree in Software Engineering in 2017 and an M.Sc. in Advanced Computer Studies in 2018 both at Liverpool John Moore's University respectively. He is currently SEND Graduate PhD Researcher in the School of Computing and Mathematics at Keele University and is a member of the Software and Systems Engineering research group. His research interest is in software engineering, AI (Machine Learning, Deep Learning), Internet of Things, Industrial IoT and Digital Twins.

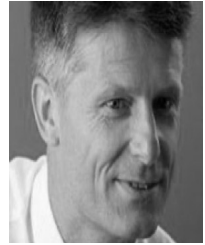

CHARLES DAY gained an undergraduate degree in physics and computer science. After graduation, he worked in the IT industry supporting manufacturing companies in the telecoms and agricultural sectors of the economy for a number of years. He returned to academia in the 1990s to complete an MSc and then a PhD (Neural Networks). After spells as a post-doctoral researcher at the EC Joint Research Centre in Italy and also in the MacKay Institute for Communication & Neuroscience at Keele, he became a Computer Science lecturer at Keele in 2001. Has special interests in computational modelling, echo state networks and data mining of large data sets.

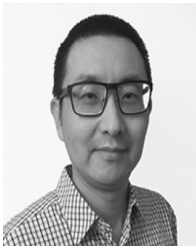

ZHONG FAN is a Professor and Academic Director of SEND at Keele University. Before that, he was Chief Research Fellow with Toshiba Research Europe, Bristol, U.K., leading research on IoT, smart grid, data analytics, and 5G communications. He was a Research Fellow with Cambridge University, a Lecturer with Birmingham University, and a Researcher with Marconi Laboratories, Cambridge. He also received a BT Short-Term Fellowship for his work at BT Laboratories. He received the B.S. and M.S. degrees from Tsinghua University, China, and the PhD degree from Durham University, U.K. His research interests are smart energy, IoT, and machine learning applications.

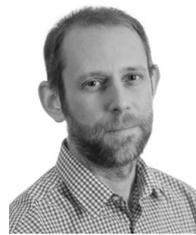

CHRIS BARLOW has been a director of SMEs for over 20 years, focused on delivering innovative solutions that improve the operations of our manufacturing, utility and broadcasting clients. Currently, as technical director of Astec IT Solutions since 2005, he is focused on growing the business to be one of the leading system integrators in the UK, providing thought-leadership around digital transformation and the impact and opportunity IIoT presents to our customers. Areas of expertise include IIoT, MES/MOM, SCADA systems, IT platforms, analytics, reporting and virtualisation.